\documentclass[twocolumn,aps,prl,showpacs]{revtex4}
\usepackage{graphicx}
\usepackage{amsfonts}
\usepackage{textcomp}

\begin{document}
\title{Realizing Universal Edge Properties in Graphene Fractional Quantum Hall Liquids}

\author{Zi-Xiang Hu$^1$}
\author{R. N. Bhatt$^1$}
\author{Xin Wan$^2$}
\author{Kun Yang$^3$}
\affiliation{$^1$Department of Electrical Engineering, Princeton University, Princeton, New Jersey
08544, USA}
\affiliation{$^2$
Zhejiang Institute of Modern Physics, Zhejiang University, Hangzhou 310027, P.R.
China}
\affiliation{$^3$National High Magnetic Field Laboratory and Department of Physics, Florida State University, Tallahassee, Florida 32306, USA}

\date{\today}
\begin{abstract}
Universal chiral Luttinger liquid behavior has been predicted for fractional quantum Hall edge states,
but so far has not been observed experimentally in semiconductor-based two-dimensional electron gases.
One likely cause of this absence of universality is the generic occurrence of edge reconstruction in such systems,
which is the result of a competition between confinement potential and Coulomb repulsion. We show that due to a completely different mechanism of confinement,
edge reconstruction can be avoided in graphene, which allows for the observation of the predicted universality.
\end{abstract}

\maketitle

{\it Introduction.} --
Topological states of matter often support protected gapless edge or surface excitations, which in turn provide a window to
probe the topology of the bulk phases. This is the case for integer and fractional quantum Hall (FQH) liquids, and topological
insulators and superconductors. The FQH edge states are described by chiral Luttinger liquid (CLL) theory~\cite{Wen}, and
predicted to exhibit certain {\em universal} low-energy properties like electron tunneling exponents, for many bulk filling
factors including the celebrated Laughlin sequence~\cite{Wen}. Such universality, however, has {\em not} been observed in
FQH liquids formed in semiconductor-based two-dimensional electrons gases (2DEGs)~\cite{grayson}. This is clearly one of the most
significant long-standing puzzles in the field of quantum Hall effect. One likely cause of this discrepancy is edge
reconstruction~\cite{chamon,xinprl}, which induces additional {\em non-chiral} edge modes that are not tied to the bulk
topology; these additional modes ruin the predicted universality~\cite{chamon,yangprl03, chang03, wan05}. Edge reconstruction
is a consequence of competition between confinement potential that holds the electrons in the interior of the sample, and Coulomb
repulsion that tends to spread out the electron density. Detailed numerical studies~\cite{xinprl, xinprb03,jain} show that due to the
electrostatic configuration of semiconductor 2DEG, edge reconstruction occurs generically in the FQH regime. This suggests
it is unlikely that one can observe the predicted universality in these systems, without carefully designing their electrostatic structure.

Recently, graphene has emerged as a brand new 2DEG system, in which FQH effects have been observed~\cite{Kirill,XDu,dean}.
Its many fascinating properties due to the Dirac nature of the electrons notwithstanding, we point out that the mechanism
for electron confinement is very different between graphene and semiconductor 2DEGs. In the latter electrons are provided
by dopants, which are positively charged (so that the system is overall charge neutral), and provide the dominant source
of confinement potential. Since these dopants are placed hundreds or thousands of angstroms away from the 2DEG, the confining
potential is not strong enough to balance the Coulomb repulsion at the edge, and prevent edge reconstruction~\cite{xinprl,xinprb03, jain}.
In graphene, on the other hand, electrons come from (properly biased) metallic gates. The charge neutrality is reflected by
the fact that every electron carries an opposite image charge due to the metallic gate; as a result the electron-electron
interaction is of dipole-diploe type at long distance, and no additional neutralizing back ground charge is present (or needed).
Thus the one-body confinement potential is provided instead by the presence of graphene boundaries.

In this work we study the combined effects of the Dirac nature of electrons and the dipole nature of interaction, and show
that there is an experimentally accessible parameter window in which edge reconstruction can be {\em avoided}. This points
to the possibility of realizing universal CLL physics at graphene quantum Hall edges, and resolving a long-standing puzzle in the field.

{\it Model.} -- In this Letter we study $\nu=1/3$, at which the first
FQH state was observed in both semiconductor
2DEGs~\cite{Tsui} and in single-layer graphene~\cite{Kirill,XDu}. In a semiconductor
2DEG, the 1/3-filled lowest Landau level (LL) originates from a single
band of nonrelativistic electrons. In graphene, however, the $0th$ LL
has a 4-fold spin and valley degeneracy with equal number of states
coming from the conduction and valence bands that form two Dirac cones
in a Brillouin zone. Hence the $\nu=0$ state corresponds to a
half-filled $0th$ LL and is already nontrivial as there are multiple
ways to occupy the LL. At zeroth order, Coulomb interaction is spin-
and valley-independent, giving rise to an internal SU(4) symmetry,
which is spontaneously broken due to the exchange effect that gives
rise to quantum Hall ferromagnetism~\cite{yangreview}. Various
symmetry-breaking perturbations lift the SU(4) degeneracy and select
the occupied states [or a specific direction in the
  SU(4)-order-parameter space]. The two obvious possibilities for the
half-filled $0th$ LL are fully-spin-polarized valley-singlet, or
fully-valley-polarized spin-singlet. Experimentally, an insulating
state was observed at $\nu=0$ at high fields~\cite{graphene_ong};
this is most naturally understood as the formation of the latter, as a
spin-polarized valley-singlet supports gapless edge states and is
conductive~\cite{abanin07,breyfertigedge,shimshoni}. We will thus
assume that the 1/3 state is built on an inert fully-valley-polarized
spin-singlet $\nu=0$ state (with {\em no} edge states). This means
that the electrons condensing into the 1/3 FQH liquid {\em no longer}
has a valley degree of freedom, but their spins remain active. In our
numerical study below, we will first assume the electron spins are
also frozen out by their Zeeman coupling to the external magnetic
field, and then release this constraint to study their possible roles
both in the bulk and at the edge.

\begin{figure}
\includegraphics[width=8.6cm,height=5.5cm]{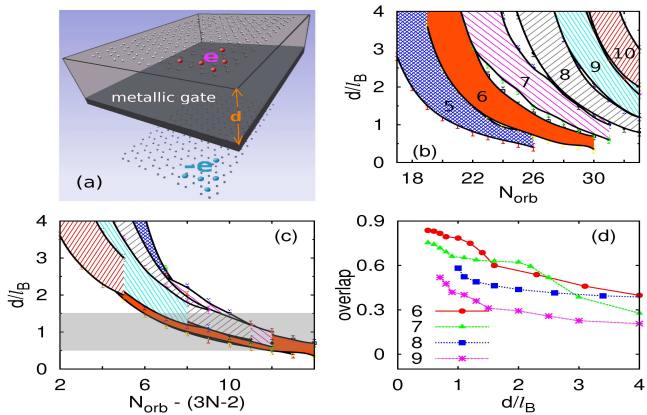}
\caption{\label{setup} (Color online) (a)Illustration of the system. A metallic gate is placed at a distance $d$ away from a single-layer graphene.  Each electron in the graphene has a positive image
  charge at a distance $2d$ below the graphene layer.(b) Regions of the
  $d$-$N_{orb}$ parameter space in which the global ground state has
  the same total angular momentum $M$ as the Laughlin state $M_0 =
  3N(N-1)/2$, for $N=5$-10 electrons. (c) Collapse of regions for
  different system sizes when (b) is replotted against $N_{edge} = N_{orb} -
  (3N-2)$. The Laughlin-like regions for $d\sim l_B$
  (roughly $0.5 \lesssim d/l_B \lesssim 1.5$, shaded area) are not very sensitive to
  system size, hence a promising window for the observation of
  universal edge physics. (d) The overlaps between the ground-state and
Laughlin wavefunction as a function of $d$ in
the Laughlin phase (along selected paths).}
\end{figure}

We consider a graphene layer situated at a distance $d$ from a
metallic gate as illustrated in Fig.~\ref{setup} (a).
The electrostatic configuration has some similarity to that of ordinary 2DEG, with the image charge playing a role similar to that of the back ground dopant
charge (see, {\em e.g.}, Fig.1 of Ref.~\onlinecite{xinprb03}). As we will see later, the dimensionless ratio $d/l_B$ ($l_B = \sqrt{\hbar c/eB}$ is the magnetic length)
controls the stability of the Laughlin state in both cases. However here the image charge moves with the electrons (and thus enters the two-body interaction) while
the background charge is inert and provides the dominant part of one-body confining potential.
It also bears some resemblance to
that in Ref.~\onlinecite{zpapic}, in which
the image charges from the dielectric media modify the form of Coulomb interaction.
Here the effective interaction between
electrons contains their direct interaction in the graphene layer and
their interaction with image charge on the other side of the metallic
gate at a distance $2d$ away:
\begin{equation}
V(r) = \frac{e^2}{r} - \frac{e^2}{\sqrt{r^2+(2d)^2}}.
\label{eq:diploeinteraction}
\end{equation}
$V(r)$ crosses over from $1/r$ at short distance to $1/r^3$ at long distance around $r \sim d$, and the  significant
pseudopotentials ($V_m$) are those with $m\lesssim d/l_B$ .

The one-body confining potential in graphene comes exclusively from the existence of a boundary where the electron wave function must vanish. In a semi-infinite graphene sheet with an
edge at $x=0$ in a magnetic field, the Dirac electron spectrum
satisfies $D_\mu(-\sqrt{2}x_c) = 0 $,
where $D_\mu(x)$ is the parabolic cylinder function, $x_c = kl_B$, $k$ is wave vector parallel to the edge and
$\mu = \epsilon^2/2$; $\epsilon$ is energy in units of $\hbar v_F/l_B$. The solutions give the energies for
the LLs in the presence of such a boundary~\cite{edgepotential}.  In the bulk, the solutions are $\mu =
0, 1, 2, \cdots$, hence the energy for the $n$th LL is $\epsilon_n
=\sqrt{2n}$.  Close to the boundary, the LLs are no longer flat. For a
sufficiently large quantum Hall droplet occupying $N_{orb}$ orbitals
on a disk with the edge at a radius $r_c = \sqrt{2N_{orb}}l_B$, we can
recast the Dirac solution as
\begin{equation}
 \label{diraceq}
D_\mu[-\sqrt{2}(x_c-r_c/l_B)] = 0.
\end{equation}
 We note
that the LL energies solved from the Dirac equation are in units of $\hbar v_F/l_B$, while the electron Coulomb interaction is in units of $e^2/l_B$ (the choice of energy unit from now on). In analogy to the definition of the
fine-structure constant $\alpha = e^2/(\hbar c) \sim
1/137$, we define the fine-structure constant for graphene $ \alpha_g  = \frac{e^2}{\hbar v_F} = \frac{e^2/l_B}{\hbar v_F/l_B} \sim 2.2$.
The largeness of the constant is due to the speed of light being much
larger than the graphene Fermi velocity $v_F \sim 10^6 m/s$.  We
feed the solution of the lowest LL (LLL) energy at $x_c = r_c/l_B - \sqrt{2m}$, scaled by $1/\alpha_g$, as the edge potential on the $m$'th orbital $U_m$ in the disk geometry.
This results in the Hamiltonian
\begin{equation}
\label{Hamiltonian}
 H = \frac{1}{2}\sum_{mnl}V_{mn}^l c_{m+l}^+c_n^+c_{n+l}c_m +
     \sum_m U_m c_m^+ c_m,
\end{equation}
where $c_m^+$ is the creation operator for an electron with angular
momentum $m$ in the LLL, $V_{mn}^l$ is the interaction matrix element calculated from Eq. (\ref{eq:diploeinteraction}).

{\it Stability of Laughlin state for spin-polarized electrons.} --
The
Laughlin state is the exact zero-energy ground state with
a total angular momentum $M = M_0 = 3N(N-1)/2$ at $\nu=1/3$ for the
hard-core potential with $V_1 > 0$ and $V_{m>1} = 0$, in the absence of one-body (or confinement) potential. Positive $V_{m>1}$, or the longer-range components of the
interaction, tend to spread out the electron density (against confining potential) and increase $M$ for the ground state; this is the driving force of edge reconstruction instability.
Compared to that
for a realistic GaAs/AlGaAs 2DEG system with long-range Coulomb
interaction~\cite{xinprl,xinprb03} (hereafter referred to as the GaAs
case), graphene has a shorter-range interaction and a
different source of edge confinement potential. However, they share some common
ingredients, e.g., a control parameter $d$ that dictates the electrostatic configuration.
Therefore, we expect that only for a finite range of
$d\sim l_B$ can the Laughlin phase be stabilized, as in the case of GaAs~\cite{xinprl}. This window can be determined by examining
the total angular momentum $M$ of the ground state and its overlap
with the Laughlin state.

In Fig.~\ref{setup}(b) we map out in the $d$-$N_{orb}$ plane the
regions in which the Hamiltonian has a global ground state at $M = M_0
= 3N(N-1)/2$ (i.e., the same as Laughlin
state), for 5-10 electrons. The Laughlin regions first rapidly shift to smaller $d$ as $N_{orb}$ increases, and
then level off for $d\sim l_B$. Correspondingly, for large $d$ the Laughlin-like state exists for a very small range of $N_{orb}$, while this range gets significantly larger for smaller $d$.
The high sensitivity on $N_{orb}$ for larger $d$ indicates that the ground state with $M = M_0
= 3N(N-1)/2$ suffers from strong finite-size effects, related to the fact that in such finite size system the edge confining potential can also impact the bulk region (near the center of the disc) strongly.
This point is supported by Fig.~\ref{setup}(c), where we replot the regions by shifting $N_{orb}$ to
$N_{edge} = N_{orb} - (3N-2)$, where $3N-2$ is the least $N_{orb}$ needed to support the Laughlin state, in an attempt to separate out the
sensitivity to the system size. We find in such a plot
the Laughlin regions are dependent on system size for larger $d$, while they fall on top of each other and show
very weak size dependence for (shaded region)
\begin{equation}
0.5 l_B \lesssim  d \lesssim 1.5 l_B,
\label{eq:stabilitywindow}
\end{equation}
indicating genuine stability of the Laughlin state in this region~\cite{note}. Our conclusion is further supported by calculating the overlap between the numerical ground states with the Laughlin state itself.
As is clearly visible in Fig. \ref{setup}(d), the overlap is very small for large $d$, but rapidly increases toward 1 as one approaches the stability window of Eq. (\ref{eq:stabilitywindow}).

The stability window of the Laughlin state, Eq. (\ref{eq:stabilitywindow}), is quite similar to that of GaAs case modeled in Refs. \onlinecite{xinprl,xinprb03,jain}, where a closely related parameter $d$ characterizes the distance between the 2DEG and the background charge due to the ions. This is not an accident; the stability window is dictated by the very similar electrostatic configurations of both cases. As discussed in detail in Refs. \onlinecite{xinprb03} (see in particular its Fig. 1), the 2DEG and its corresponding neutralizing positive ion charge form a capacitor; the fringe field near the edge is the driving force of edge reconstruction instability. Once $d$ becomes large compared to $l_B$ which is the size of the LLL wave function, such an instability occurs. Here the situation is almost identical, with the capacitor formed by graphene 2DEG and the metallic gate. In GaAs $d$ is typically of order 1000 {\AA} or more than $10l_B$ (essential for high mobility), thus edge reconstruction is essentially unavoidable. In graphene on boron-nitride, however, $d$ can be as small as a few nanometers or a small fraction of $l_B$\cite{dean,young}; as a result the stability window (\ref{eq:stabilitywindow}) is within reach. We predict that in such a window edge reconstruction and other related instabilities can be avoided, resulting in a {\em single} chiral edge mode for the 1/3 FQH state. In this window single electron tunneling\cite{grayson,chang03} will exhibit a non-linear I-V characteristic $I\propto V^\alpha$ with a {\em universal} exponent $\alpha=3$, as predicted by theory\cite{Wen}.

\begin{figure}
\includegraphics[width=8.8cm,height=3.5cm]{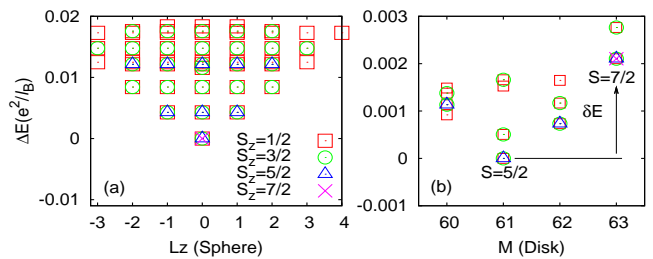}
\caption{\label{spinspectrum} (Color online) Low-energy spectrum for 7
  electrons on (a) a sphere ($N_\phi=18$, $d=1.0l_B$) and (b) a disk ($N_{orb}=26$
  and $d=1.8l_B$).  The ground state is spin fully-polarized on the sphere
  and partially-polarized on the disk. $\delta E$ is the energy gap
  between the ground state and the lowest energy spin fully-polarized state. }
\end{figure}

{\it Spin partially-polarized ground states.} -- We now switch on the
spin degrees of freedom. Earlier numerical studies on torus ~\cite{zhang} and sphere~\cite{ed87} indicate that electrons at $\nu=1/3$ are fully
spin-polarized with Coulomb interaction, even in the {\em absence} of Zeeman splitting.
Our numerical calculation confirms that a system of 7 electrons on a sphere with dipolar interaction has a ground state with total spin $S
= 7/2$, as shown in Fig.~\ref{spinspectrum}(a). These results clearly suggest that we have a spin fully-polarized ferromagnet in the {\em bulk}. The situation,
however, can be more complicated at the edge. It was shown ~\cite{Karlhede,Nakajima} that the $\nu = 1$ quantum Hall ferromagnet can support a spin texture at the edge. In the following we show this can also happen for the 1/3 edge when the Zeeman splitting is sufficiently small, and calculate the magnetic field needed to fully polarize the edge.

\begin{figure}
\includegraphics[width=9cm,height=5.5cm]{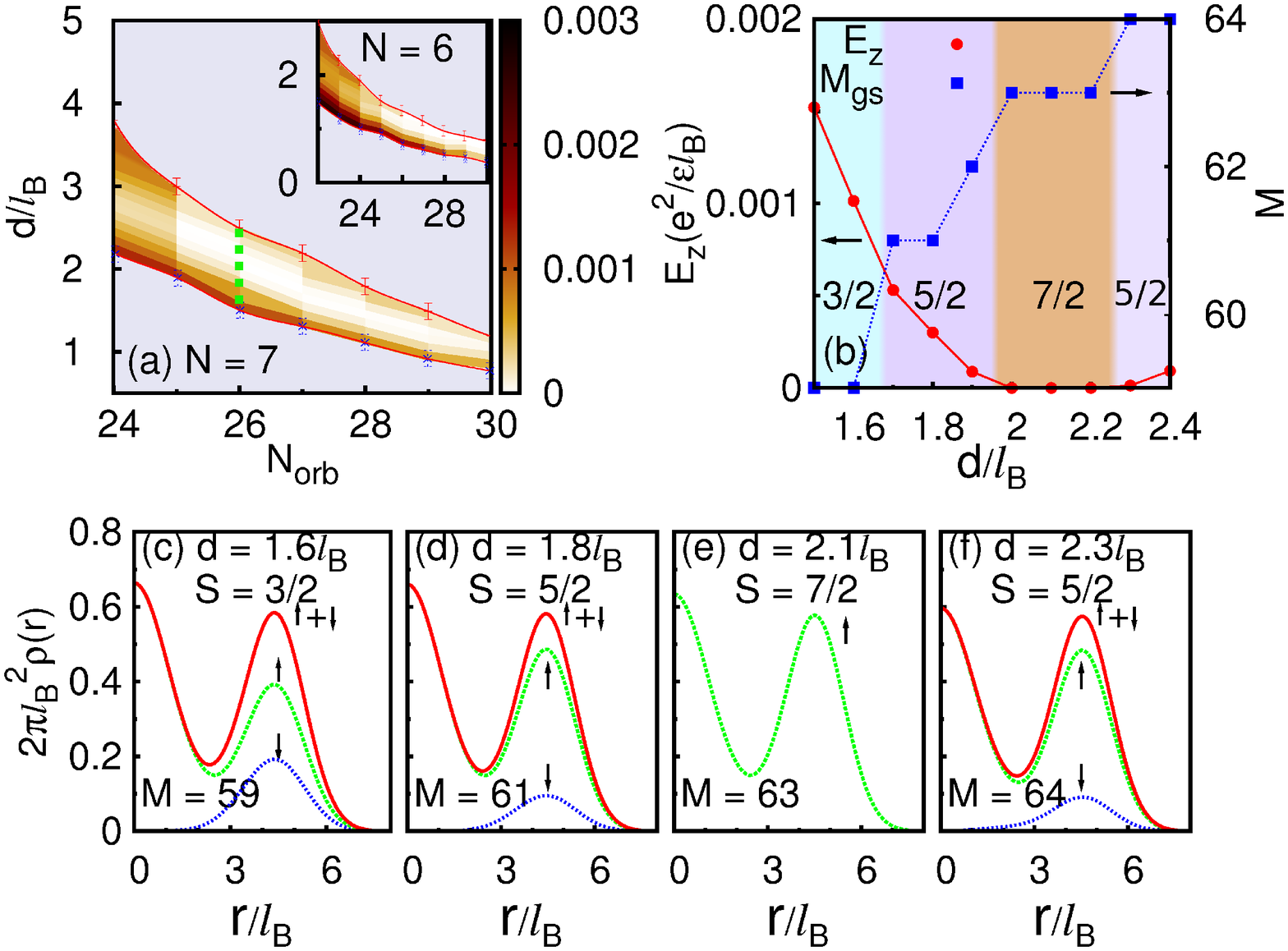}
\caption{\label{phasespin} (Color online) (a) Zeeman energy necessary
 to fully-polarize the ground state of 7 (6 for inset) spinful electrons in the
  $d$-$N_{orb}$ parameter space.
  (b) The Zeeman energy
  and the ground state total angular momentum for the 7 electrons
  along the (green) dashed line in (a) (i.e., $N_{orb} = 26$).  (c),
  (d), (e) and (f) plot the spin-resolved density profiles for 4
  selected ground states in the 4 different spin regions in
  (b). }
\end{figure}

We choose 7 electrons in 26 orbitals at $d=1.8l_B$ as an example, for which
the ground state of spin-polarized electrons is inside the Laughlin-like region as
shown in Fig.~\ref{setup}. Fig.~\ref{spinspectrum}(b) plots the low-lying energy
spectrum of the system, in the absence of Zeeman splitting. The global ground
state now has $S=5/2$, indicating the ground state is not spin fully-polarized.
To fully polarize the
ground state, we need to apply a finite Zeeman energy, which exceeds the
energy difference per electron between the two states: $E_{z} \ge \delta
E / N = (E_{polarized} - E_{gs})/N$.
Fig.~\ref{phasespin}(a) plots in color the minimum Zeeman energy
required to polarize the ground state to be Laughlin-like for
systems of 6 and 7 electrons.
In both cases we find the required
$E_z\sim 0.003 e^2/ l_B$ with essentially no size dependence; this indicates both bulk {\em and} edge electron spins are fully polarized for $B \gtrsim 2$ Tesla (with electron spin $g$ factor 2). In these cases results of the previous section remain valid.

To reveal the nature of the
spin partially-polarized states (which can be stable at sufficiently low magnetic fields), we fix $N_{orb} = 26$ and vary $d$
inside the Laughlin-like region, indicated by the vertical (green)
dashed line in Fig.~\ref{phasespin}(a).  As shown in
Fig.~\ref{phasespin}(b), the total spin $S$ and the angular momentum
$M_{gs}$ for the ground state both vary with $d$, indicating the interplay between edge confinement and spin structure.
Fig.~\ref{phasespin}(c)-(f) show the
spin-resolved density profiles of 4 ground states in the corresponding
regions distinguished by the ground state spin $S$ in
Fig.~\ref{phasespin}(b).
The total density profile for these cases are
almost identical. However, when the ground state is
partially polarized, the minority-spin electron(s) (2 electrons for $S =
3/2$ and 1 for $S = 5/2$) form a ring-like puddle along the edge;
there is no electron with the minority spin near the center. This clearly indicates the reduction of polarization is due to spin texture formation at the edge, similar to that of $\nu=1$\cite{Karlhede,Nakajima}.
These edge spin textures, while eliminated by moderately high magnetic fields and unfavorable for the universal edge physics sought in this paper, are interesting in their own right and deserve further study, in light of the current experimental\cite{Nakajima,barak} and theoretical\cite{Barlas} interest in spin physics at quantum Hall edges.

In conclusion, we have demonstrated through detailed numerical studies that edge reconstruction can be avoided in an experimentally accessible parameter window for the $\nu=1/3$ graphene fractional quantum Hall liquid, allowing for observation of {\em universal} chiral Luttinger liquid behavior. It can also support interesting edge spin textures that can be probed experimentally.

This work was supported by DOE grant No. DE-SC0002140 (Z.X.H.,
R.N.B. and K.Y.) and the 973 Program under Project No. 2009CB929100
(X.W.).


\begin{thebibliography}{99}
\bibitem{note} Further increasing $N_{orb}$ will, of course, eventually lead to expansion of area of the electron liquid, which may condense into other fractional quantum Hall states with lower filling factors. However the robustness of the 1/3 Laughlin state over a substantial range of $N_{orb}$ is clear indication of its stability.
\end{thebibliography}
\end{document}